\documentclass{elsart}

\usepackage{natbib}

\usepackage{epsfig}

\usepackage{amssymb}

\begin{document}

\begin{frontmatter}



\title{A Well-Behaved $f(R)$ Gravity Model\\
in Planetary Motions
}


\author{Robab Hashemi 
}
\address{Department of Physics,
University of Guilan, P. O. Box: 41335-1914, Rasht, Iran}

\author{Reza Saffari\corauthref{cor}
}
\address{Department of Physics,
University of Guilan, P. O. Box: 41335-1914, Rasht, Iran}
\corauth[cor]{Corresponding author}
\ead{rsk@guilan.ac.ir}


\begin{abstract}

In this paper we consider asymptotic behavior of a hybrid action of
$f(R)$ gravity model which proposed by \citet{saf08}, in the Solar
system scale, which can explain Pioneer anomalous acceleration. We
use the resultant weak field gravitational potential which comes
from the hybrid action to test its impacts on the Solar system
dynamics, by comparing theoretical precession of perihelion of a
test particle, $\dot{\varpi}$, with corrections to the standard
Newtonian - Einstenian precessions of perihelia of some planets,
which recently estimated by \citet{pit1,pit2,pit3,pit4}. Here we
show that the asymptotic behavior of hybrid action, are in more
accordance with observation relative to the other modifications such
as power law and logarithmic corrections \citep{irio}. We also show
that an extra additional lensing of the prediction of General
Relativity is reproduced. Finally we consider the stability
condition of planetary orbits in presence of the hybrid action.

\end{abstract}


\end{frontmatter}

\parindent=0.5 cm

\section{Introduction}
\label{intro}

To accommodate some recently observed phenomena
\citep{riess,perl,cole,teg,sper03,sper07,clowe,ander98,ander02},
occurring at very different scales ranging from Solar system to
cosmological distances, which at present, have not yet found fully
satisfactorily explanations in terms of conventional physics,
gravitation or not, \citep{Paramos,Bertolami,Lammerzahl}, we
considering modifications of the standard laws of gravity at
different scales; As well as done in recent years. These anomalous
effects are: The recent data coming from the luminosity distance of
SuperNovae Ia (SNeIa)\citep{riess}, wide galaxy surveys \citep{cole}
and the anisotropy of cosmic microwave background radiation
\citep{sper03,sper07} suggest that about $70$ percent of the energy
density of the present universe is composed by dark energy,
responsible for an accelerating expansion. Further more in small
scales, the data coming from galactic rotation curves of spiral
galaxies \citep{clowe} can not be explained on the basis of
Newtonian gravity or General Relativity (GR), if one does not
introduce invisible dark matter. In Solar system there is an
unexplained acceleration $a_p=(8.47\pm 1.33)\times10^{-10} ms^{-2}$
approximately directed towards the Sun affecting the Pioneer $10/11$
spacecrafts after they passed $20 AU$ threshold
\citep{ander98,ander02}. In order to reconcile theoretical models
with observations, existence of dark matter is postulated in Solar
system \citep{Nieto}. The main problem of dark energy and dark
matter is understanding their nature, since they are introduced as
ad hoc gravity sources in a well define model of gravity. Therefore
we find another possibilities, like various modification of gravity.
In this paper we are concerned with $f(R)$ gravity theory, where the
gravitational lagrangian depends on a function of the scalar
curvature $R$ \citep{Nojiri7,Capozziello,Faraoni}. These theories
are also referred to as 'extended theories of gravity', since they
naturally generalize GR: in fact, when $f(R)=R$ the action reduces
to the usual Einstein-Hilbert action, and Einstein's theory is
obtained. These theories can be studied in the metric formalism,
where action is varied with respect to metric tensor and in Palatini
formalism, where action is varied with respect to metric and affine
connections, which are supposed to be independent from one to
another. In general, two approaches are not equivalent: the
solutions of Palatini field equations are a subset on metric field
equations \citep{Magnano}. Actually, $f(R)$ theories provide
cosmologically viable models, where both the inflation phase and the
accelerating expansion are reported \citep{cruz,Nojiri}.
Furthermore, they have been used to explain the rotation curves of
galaxies without need for dark matter \citep{Frigerio}. However,
because of the excellent agreement of GR with Solar system and
binary pulsar observations, every theory that aims at explaining
galaxies dynamics and the accelerating expansion of the Universe,
should reproduce GR at the Solar system scale, in a suitable weak
field limit. Without going into the details, we want to test the
model of $f(R)$ gravity and using perturbative approach, we focus on
the impacts which modification of the gravitational field due to
nonlinearity of the gravitational lagrangian, have on Solar system
dynamics. We compare the theoretical prediction for the precession
of the longitude of the pericenture of a test particle with the
correction to the standard Newtonian-Einsteinian precession of
perihelia of some planets recently estimated by
Pitjeva.\citep{pit1,pit2}. At first, we derive an explicit
expression of the secular, averaged over one orbital revolution
perihelion precession, induced by anomalous acceleration of
asymptotic behavior of hybrid action on the orbits of Solar system
planets. Then we will compare obtained results in small eccentricity
approximation, with the latest estimated corrections of the
perihelion rates. It is important to note that the corrections to
the perihelion rates determined in \citep{pit3,pit4}, are
phenomenologically estimated quantities of a global, least-square
solution in which only Newtonian and Einstenian dynamics was
modeled; no exotic dynamical terms were included in the fit. Thus in
our opinion such phenomenological corrections can genuinely be used
to get information on a hypothetic, un-modeled force. Also, central
to the development of the good modified theory of gravity that can
describe the phenomenology usually ascribed to dark matter in the
analysis of the amount of deflection of light implied by the theory.
Then we obtain deflection of light for the hybrid model and compare
it with the prediction of GR. This paper is organized as follows: in
section \ref{fofr}, we briefly review the theoretical formalism of
$f(R)$ gravity, in section \ref{proposal}, we outline a general
approach to the perturbations of the gravitational field of GR, in
section \ref{asymp}, we introduce asymptotic behavior of hybrid
$f(R)$ model and we compare theoretical predictions with
observations and compare our results with some models that tested in
Solar system, in section \ref{def}, we calculate deflection of
light, in section \ref{stab} we consider stability of circular
orbits and finally, conclusions are in section \ref{con}.

\section{$f(R)$ gravity}
\label{fofr}

Here we introduce the field equation of $f(R)$ gravity
\citep{Sotiriou}, we start from  modified Einstein-Hilbert action
\begin{equation}
S=\frac{1}{2k}\int d^4x\sqrt{-g}f(R)+S_m. \label{start}
\end{equation}
The gravitational part of the Lagrangian, $f(R)$ is represented by a
function of the scalar curvature, $R$. $S_m$ is the matter part of
the action. In metric formalism affine connections, $\Gamma$'s are
supposed to be Levi-Civita connections of $g$ and consequently, the
scalar curvature $R$ has to be intended as
\begin{equation}
R\equiv R(g)=g^{\alpha\beta}R_{\alpha\beta}(g).\label{metric}
\end{equation}
On the contrary in the Palatini formalism metric $g$ and affine
connections are supposed to be independent, so that the scalar
curvature $R$ has to be intended as
\begin{equation}
R\equiv
R(\Gamma)=g^{\alpha\beta}R_{\alpha\beta}(\Gamma),\label{pala}
\end{equation}
where $R_{\alpha\beta}(\Gamma)$ is the Ricci-like tensor of the
connections $\Gamma$. We follow the metric formalism that in this
regime, Eq. (\ref{start}), is varied with respect to the metric
$g_{\mu\nu}$, and obtain the field equation of motion
\begin{equation}
F(R)R_{\mu\nu}-\nabla_{\mu}\nabla_{\nu}F(R)-kT_{\mu\nu}=
g_{\mu\nu}(\frac{1}{2}f(R)-\nabla_{\alpha}\nabla^{\alpha}F(R)),
\label{field}
\end{equation}
where $T_{\mu\nu}$ is the standard minimally coupled matter energy-
momentum tensor and $F=df/dR$. Contraction of the field equation Eq.
(\ref{field}), with metric tensor leads to scalar equation
\begin{equation}
3\nabla_{\alpha}\nabla^{\alpha} F(R)+F(R)R-2f(R)=\frac{8 \pi
G}{c^4}T,\label{trace}
\end{equation}
where $T$ is the trace of the energy-momentum tensor. Eq.
(\ref{trace}) is a differential equation for the scalar curvature
$R$. In order to compare the prediction of $f(R)$ gravity with Solar
system data, we have to consider the solutions of the field equation
Eq. (\ref{field}) supplemented by the constraints Eq. (\ref{trace})
in vacuum, since tests are based on the observations of the dynamics
of the planets in the gravitational field of the Sun. We notice that
if $R=constant$ we obtain the Palatini case: so for a given $f(R)$
function in vacuum, solutions of the field equation of Palatini
$f(R)$ gravity are a subset of the field equation of metric $f(R)$
gravity. \citep{Multamaki}

\section{The proposal method to test modified $f(R)$ gravity model}
\label{proposal}

In this section we outline the general procedure that we are going
to apply to the solution of $f(R)$ gravity, In order to compare the
prediction of these gravity models with the existing data. In weak
field approximation of metric and spherically symmetrical condition
\begin{equation}
ds^2=-B(r)dt^2+A(r)dr^2+r^2\sin^2\theta d\varphi^2+r^2
d\theta^2,\label{metrici}
\end{equation}
the gravitational (scalar) potential $\phi(r)$ is read from the
$B(r)$ function
\begin{equation}
B(r)=1+\frac{2\phi(r)}{c^2}.\label{coaf}
\end{equation}
We expect a gravitational potential in the form of
\begin{equation}
\phi(r)=\phi^{N}(r)+\triangle \phi(r),\label{pert}
\end{equation}
where $\phi^{N}(r)=-GM/r$ is the Newtonian potential of a point-like
mass $M$ in center, and $\triangle \phi(r)\ll\phi^{N}(r)$ is a
correction vanishing for $f(R)\rightarrow R$. Now if we have a
correction for Newtonian potential, we can calculate the perturbing
acceleration and then we will be able to calculate its effect on
planetary motions, within standard perturbative schemes \citep{Roy}.
Gauss equation for entirely radial perturbing acceleration $A_{r}$
is
\begin{equation}
\frac{d
\varpi}{dt}=-\frac{\sqrt{1-e^2}}{nae}A_{r}\cos\varphi.\label{Guas}
\end{equation}
After being evaluated onto the unperturbed Keplerian ellipse, the
acceleration $A_{r}$ must be inserted into Eq. (\ref{Guas}); then,
the average over one orbital period $P$ must be performed. To this
end, we use following equations; where $n=\sqrt{{GM}/{a^3}}$, is the
Keplerian mean motion, $a$ is the planets semi major axis, $e$ is
the eccentricity and $\varphi$ is the true anomaly. \citep{Roy}
\begin{equation}
\cos \varphi=\frac{\cos E-e}{1-e\cos E},\label{help1}
\end{equation}
\begin{equation}
r=a(1-e\cos E),\label{help2}
\end{equation}
\begin{equation}
dt=\frac{1-e\cos E}{n}dE.\label{help3}
\end{equation}
In fact, what we aim at, is evaluating the perturbations induced on
longitudes of perihelia by the corrections to the gravitational
field due to $f(R)$ gravity. In order to compare them with the
latest observation we use table 1 and 2 of \citet{pit1,pit2}, that
recently processed almost one century of data of different types for
major bodies of the Solar System in the effort of continuously
improving the $EPM 2004/EPM 2006$ planetary ephemerides,
\citep{pit3}. Among other things, she also estimated corrections to
the secular rates of the longitudes of perihelia of some planets of
the Solar system as fit-far parameter of a global solution in which
she contrasted, in a least-square way, the observations to their
predicted values computed with a complete set of dynamical force
models including all the known Newtonian and Eisteinian features of
motion. As a consequence, any un-modeled exotic force present in
nature is; in principle, entirely accounted for by the obtained
apsidal extra precession $\Delta\dot{\varpi}$. See table
\ref{observations} for inner planets and table \ref{observations2}
for outer planets of Solar system.

Now let us briefly outline how we are going to put $f(R)$ gravity on
the test. We follow Iorio's regime for testing modified gravity
models, \citep{irio} and take the ratios of observational
$\Delta\dot{\varpi}$ for different pairs of planets $A$ and $B$
\begin{equation}
\Omega_{AB}=\frac{\Delta\dot{\varpi}_{A}}{\Delta\dot{\varpi}_{B}},\label{obs}
\end{equation}
and compare it with the predicted ratios of modified model of
gravity $\xi_{AB}$. If we have $\Psi_{AB}=|\Omega_{AB}-\xi_{AB}|=0$
within the errors, the $f(R)$ gravity model examined, can still be
considered compatible with data, otherwise it is seriously
challenged.

\section{Asymptotic behavior of the hybrid action}
\label{asymp}

\begin{figure}
\begin{center}
\includegraphics*[width=12cm,angle=0]{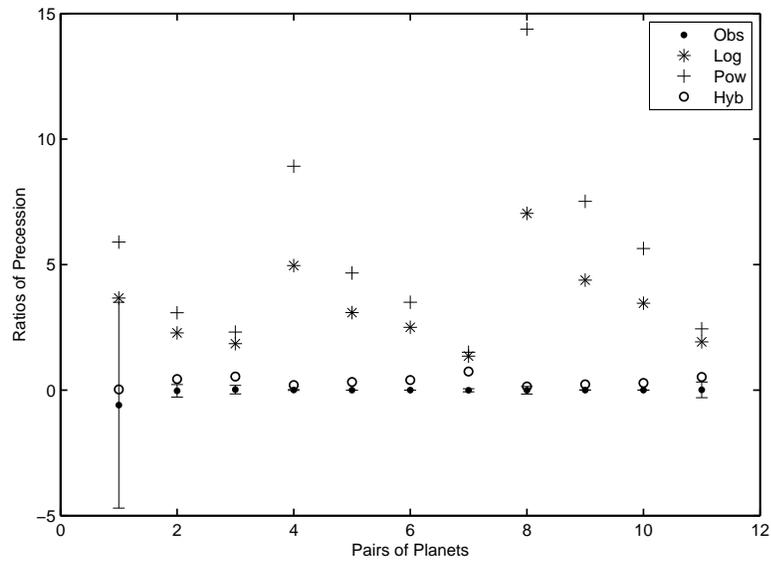}
\end{center}
\caption{This figure shows accordance of observational precession
with hybrid action theoretical results against power-low and
logarithmic corrections.} \label{fig1}
\end{figure}

\citet{saf08}, use the inverse solution to extract an appropriate
action for modified gravity,
\begin{equation}
f(R)=R+\Lambda+\frac{R+\Lambda}{R/R_{0}+2/\alpha}\ln\frac{R+\Lambda}{R_c},\label{genact}
\end{equation}
where $R_0={6\alpha^2}/{d^2}$ and $R_c$ is the constant of
integration, $\alpha\simeq 10^{-6}$, is a small dimensionless
constant and $d\simeq 10 kpc$, is length scale in the order of
galactic size and $\Lambda$ is cosmological constant. In Solar
system, for the range of $R\gg\Lambda$ and $R/R_0\gg 2/\alpha$,
action reduces to
\begin{equation}
f(R)=R+R_0 \ln \frac{R}{R_c}.\label{main act}
\end{equation}
On the other hand, in galactic scales, for $\alpha\ll 1$ and
$R\simeq R_0\simeq \Lambda$ generic action can be written as
\begin{equation}
f(R)=(R+\Lambda)[1+\frac{\alpha}{2}\ln(\frac{R+\Lambda}{R_c})],
\end{equation}
which can justifies flat rotation curves of spiral galaxies, and for
small $\alpha$ can be written as
\begin{equation}
f(R)=\frac{(R+\Lambda)^{1+\alpha/2}}{R_c^{\alpha/2}},
\end{equation}
which reduces to $f(R)=R+\Lambda$ for $\alpha\ll 1$. In Solar system
scale, in which the range of distances are sufficiently smaller than
the length scale of the hybrid action, we obtain metric elements of
Eq. (\ref{metrici}) up to the first order terms in $\alpha$ as
\begin{eqnarray}
B(r)&=&1-\frac{2GM}{c^2r}+\frac{\alpha r}{d}\nonumber\\
A(r)&=&\frac{1}{B(r)},
\end{eqnarray}
therefor the weak field effective potential which comes from Eq.
(\ref{main act}) obtains as
\begin{equation}
\phi(r)=-\frac{GM}{r}+\frac{\alpha c^2}{2d}r,\label{main}
\end{equation}
where the first term is the Newtonian potential and the second one
is corrected term. Acceleration of a test particle in this potential
is
\begin{equation}
a=-\frac{GM}{r^2}-\frac{\alpha c^2}{2d},\label{accel}
\end{equation}
where the second term in the right hand side of this equation, is a
constant acceleration, independent of the central mass. We may
correspond this extra term to the pioneer anomalous and constrain it
with the observed value of $a_p=(8.47\pm 1.33)\times 10^{-10}
ms^{-2}$ which results in ${\alpha}/{d}\cong 10^-26m^{-1}$. In this
case, correction to the gravitational potential is
\begin{equation}
\Delta \phi(r)=\frac{\alpha c^2}{2d}r,
\end{equation}
where the acceleration of a test particle in this potential is
\begin{equation}
A_r=-\frac{\alpha c^2}{2d}.\label{radial}
\end{equation}
On using
\begin{equation}
\int^{2\pi}_{0}(\cos E-e)dE=-2\pi e,
\end{equation}
and Eqs. (\ref{Guas})-(\ref{help3}) it yields to the following
perihelioin precession
\begin{equation}
<\dot{\varpi}>=-\frac{\alpha
c^2}{2d}\frac{\sqrt{(1-e^2)a}}{\sqrt{GM}}.\label{pre}
\end{equation}
It is important to note that $\dot{\varpi}$ depends on a positive
power of the semi major axis $a$. The predicted extra precession of
Eq. (\ref{pre}) can be fruitfully compared to the correction of
usual Newtonian-Einsteinian perihelion rates of the planets of the
Solar system phenomenological estimated by
\citet{pit1,pit2,pit3,pit4}. According to the general outline of
section \ref{proposal}, we will not use one perihelion at a time for
each planet. Indeed let us consider a pair of planets $A$ and $B$,
take the ratio of prediction of Eq. (\ref{pre})
\begin{equation}
\frac{\dot{\varpi}_{A}}{\dot{\varpi}_{B}}= \sqrt \frac{(1-e^2_A)M_B
a_A}{(1-e^2_B)M_A a_B}.
\end{equation}
Now we consider a pair of planets $A$ and $B$ and take the ratio of
their estimated extra-rates of perihelia in approximation small
eccentricity; if Eq. (\ref{pre}) is responsible for them,
\begin{equation}
\Psi_{AB}=|\frac{\dot{\varpi}_{A}}{\dot{\varpi}_{B}}
-(\frac{a_A}{a_B})^\frac{1}{2}|,\label{dif}
\end{equation}
then the quantity must be compatible with zero. Table \ref{res}
shows the results of observational ratio and theoretical ratio of
planets. It is important to note that the uncertainty in $\Psi_{AB}$
has been conservatively estimated by linearity adding the individual
terms coming from the propagation of the errors in $\dot{\varpi}$
and $a$ in Eq. (\ref{dif}). In general, deviation of each function
has the form of
\begin{equation}
\varrho_{f}=\sqrt{(\sum\partial f/\partial x_i)^2(\varrho_i)^2},
\label{dev}
\end{equation}
and errors in $\Omega_{AB}$ due to $\delta a$ and
$\delta\Delta\dot{\varpi}$ have the form below
\begin{equation}
\delta\Psi_{AB}\leq
|\frac{\Delta\dot{\varpi}_A}{\Delta\dot{\varpi}_B}|(\frac{\delta\Delta\dot{\varpi}_A}
{|\Delta\dot{\varpi}_A|}+
\frac{\delta\Delta\dot{\varpi}_B}{|\Delta\dot{\varpi}_B|})
+\frac{1}{2}(\frac{a_A}{a_B})^\frac{3}{2}(\frac{\delta
a_A}{a_A}+\frac{\delta a_B}{a_B}).
\end{equation}
To see that how, this model acts in the Solar system, we can compare
it with other corrections in Solar system. To this end we use
power-law and logarithmic model in weak field approximation.
\citet{Cardone} started the Lagrangian of the form $f(R)=f_0 R^n$,
in the metric approach, to obtain solutions to describe galaxies
rotation curves without need for dark matter. They obtain modified
gravitational potential of the form
\begin{equation}
\phi(r)=-\frac{GM}{r}[1+(\frac{r}{r_c})^ \beta],\label{phicapo}
\end{equation}
where $\beta$ relates to the the power of $n$ in the modified
Lagrangian via
\begin{equation}
\beta=\frac{12n^2-7n-\sqrt{36n^4+12n^3-83n^2+50n+1}}{6n^2-4n+2}.
\end{equation}
This equation clearly leads to the radial acceleration
\begin{equation}
A_r=\frac{(\beta-1)GM}{r_c^ \beta}r^{\beta-2}
\end{equation}
It yields the following perihelion precession \citep{irio}
\begin{equation}
<\dot{\varpi}>=\frac{(\beta-1)\sqrt{M}}{2\pi r_c^
\beta}a^{\beta-\frac{3}{2}}.
\end{equation}
\citet{Cardone} applied Eq. (\ref{phicapo}) to a sample of 15 low
surfaces brightness galaxies with combined $HI$ and $H\alpha$
measurements of the rotation curve extending in the putative dark
matter dominated region. They obtained a good agreement between the
theoretical rotation curves and the data using only stellar disk and
interstellar gas when the slope n of the gravity Lagrangian is set
to the value $n=3.5$ (giving $\beta=0.817$) obtained by fitting the
SNeIa Hubble diagram with the assumed power law $f(R)$ model. For
$\beta=0.817$, results are in figure \ref{fig1}. Logarithmic
correction suggested to solving the problem of dark matter in
galaxies \citep{Van Moorsel,Fabris,Sobouti}. The potential is the
form of
\begin{equation}
V_{ln}=-\alpha GM\ln(\frac{r}{r_0}),
\end{equation}
that gives the radial extra acceleration
\begin{equation}
A_r=-\frac{\alpha GM}{r},
\end{equation}
and it gives the precession to the form of
\begin{equation}
<\dot{\varpi}>=-\alpha \sqrt{\frac{GM(1-e^2)}{a}}(\frac{-1+
\sqrt{(1-e^2)}}{e^2}).
\end{equation}
The results for this model are in figure \ref{fig1}. As a conclusion
the logarithmic and power law correction is ruled out by the present
day observations \citep{irio}. As we see in figure \ref{fig1}, the
correction to the form of this asymptotic behavior of hybrid $f(R)$,
has the minimum difference with observational data. 

Now the question is, about the properties of these models. we see
that each of these corrections, has the term like $r^n$ in it. For
power law model $n=0.817$, for logarithmic correction $n=-1$ and for
the hybrid model $n=0$. then we can debate on the power of $r^n$,
and find that, which power of $n$ has the appropriate accordance
with observational data. With primary surveys we obtain $n=5.38$.
For such potential, we can find the metric and the $f(R)$ model from
inverse solution which is in preparation.

\section{Deflection of light}
\label{def}

To see that the effect of this asymptotic hybrid $f(R)$ model on the
amount of deflection, we calculate the deflection of the orbit from
a straight line
$$\Delta \Phi = 2 |\Phi(r_0) - \Phi_{00}| - \pi,$$
where $r_0$ is the closet distance to the Sun.
$$\Delta \Phi = 2 \int_0 ^{\infty} A^{1/2} (r)
\left[ \left( \frac{r}{r_0} \right) ^2 \left( \frac{B(r_0)}{B(r)}
 - 1 \right) \right] ^ {-1/2} \frac{dr}{r} - \pi.$$
By using the values of $A(r)$ and $B(r)$, we could obtain $\Delta
\Phi$ as elliptic integrals
\begin{eqnarray}
\Delta \Phi=&&GR_{terms}-\frac{1}{2}\frac{\alpha c^2}{d}r_0[\ln
\frac{r}{r_0}+\ln\sqrt{(\frac{r}{r_0})^2-1}]\nonumber\\
& &+\frac{1}{2}\frac{\alpha c^2}{d} [r_0\ln \frac{r}{r_0}+\ln
\sqrt{(\frac{r}{r_0})^2-1}]-\frac{r_0^2}
{r+r_0}\sqrt{(\frac{r}{r_0})^2-1}].
\end{eqnarray}
We replace $r$ with $r_0$, then to first order in $MG/r_0$ the
deflection is
\begin{equation}
\Delta \Phi = \frac{4MG}{r_0}
\end{equation}
and this is the prediction of GR. Then this asymptotic behavior of
hybrid $f(R)$ model do not violate GR regimes and have no any
additional effect on deflection of light.

\section{Stability of circular orbits}
\label{stab}

Finally in the context of stability condition of orbits, we test the
obtained weak field potential with numerous equation of stability:
\begin{equation}
\frac{F_c'(r)}{F_c(r)}+\frac{3}{r}>0,
\end{equation}
in which $F_c$ is central force, we obtain $r^2<{GMd}/{3\alpha
c^2}$. Then maximum of the radius of stable circular orbit, is about
$r_s\approx 2.22\times10^{11}~km$ which is $10^3$ larger than
typical Solar system radius, in order of magnitude. Therefore all
the planets in this modified regime may have stable orbits.

\section{Comments and conclusions}
\label{con}

In this paper we have studied the secular precession of the
pericenture of a test particle in motion around a central mass $M$
whose Newtonian gravitational potential exhibits a correction which
is to form of Eq. (\ref{main}). In order to put on the test the
hypothesis that such extra forces are not zero we devised a suitable
test by taking into account the ratios of the corrections to the
secular precession of perihelia estimated by
\citet{pit1,pit2,pit3,pit4}. The results obtained, resumed by Eq.
(\ref{pre}), show that modifications of the Newtonian potentials
like this examined in this paper are compatible with the currently
available apsidal extra-precession of the Solar system planets. Of
course we have to say that this model will be ruled out if we take
the inverse ratio of precession for planets. Hence we take this
regime that, take the inner planet to outer planet. Our correction
can be used to describe the problem of dark matter in the Solar
system, if we choose the appropriate values for parameters of the
model. It gives Pioneer anomalous acceleration for the suitable
value of distance. When we use the other estimative corrections to
Newtonian-Einstenian planetary perihelion rates, maybe obtain better
compatibility for this model in the Solar system. We had shown that
this asymptotic hybrid model give the GR, deflection of light and
had no extra effect on it.
 From comparison alternative models, we could offer the generic
model of potential, and obtain it's action. This proposal model has
minimum variance, and our aim is checking its compatibility with
dark matter and dark energy data set.

Acknowledgments: We would like to thank anonymous referee for useful
comments.


\clearpage

\begin{table}
\caption{observational data for inner planets}
\begin{tabular}{llll}
\hline
 planet&Mercury&Earth&Mars\\
\hline
$\Delta\dot\varpi (arcs/cy)$   &-0.000036$\pm$.005  & -0.0002$\pm$.00004  & 0.0001$\pm$.0005 \\
$a(AU)$   &  0.387  & 1     & 1.523  \\
$e$       &  0.205  & 0.016   & 0.0934      \\
$P(yr)$   &  0.24   & 1    & 1.88  \\
\hline
\end{tabular}
\label{observations}
\end{table}

\begin{table}
\caption{observational data for outer planets}
\begin{tabular}{llll}
\hline
planet&Jupiter&Saturn&Uranus\\
\hline$
\Delta\dot\varpi (arcs/cy)$   &0.0062$\pm$.036  & -0.92$\pm$2.9  & 0.57$\pm$13 \\
$a(AU)$   &  5.203 & 9.537     & 19.191  \\
$e$       &  0.0483  & 0.0541   & 0.0471      \\
$P(yr)$   &  11.86   & 29.45    & 84.07  \\
\hline
\end{tabular}
\label{observations2}
\end{table}

\begin{table}
\caption{results, $A$ $B$ denotes the pair of planets used
$\Omega_{AB}=\frac{\Delta\dot{\varpi}_{A}}{\Delta\dot{\varpi}_{B}}$,
$\xi_{AB}=(\frac{a_A}{a_B})^\frac{1}{2}$. The perihelion extra rates
for the planets have been retrieved from \citet{pit3}, their errors
are not formal; and re scaled by a factor $10$. The uncertainties in
the semi major axis have been retrieved from table 5 of
\citet{pit1}}
\begin{tabular}{llllll}
\hline
Pairs&$A$&$B$&$\Omega_{AB}$&$\xi_{AB}$&$\Psi_{AB}$\\
\hline
1 & Mer  & Jup  & -0.6$\pm$4.1       & 0.272$\pm$ $10^{-9}$ &  0.87$\pm$4.1 \\
2 & Ear  & Jup  &-0.03$\pm$0.25      & 0.438$\pm$ $10^{-8}$ &  0.46$\pm$0.25 \\
3 & Mar  & Jup  & 0.02$\pm$0.17      & 0.541$\pm$ $10^{-9}$ & 0.52$\pm$0.17  \\
4 & Mer  & Sat  & 0.004$\pm$0.017    & 0.201$\pm$ $10^{-8}$ &  0.19$\pm$0.017 \\
5 & Ear  & Sat  & 0.0002$\pm$0.0011  & 0.323$\pm$ $10^{-10}$&  0.32$\pm$0.0011 \\
6 & Mar  & Sat  & -0.0001$\pm$0.0009 & 0.399$\pm$ $10^{-8}$ &  0.399$\pm$0.0009 \\
7 & Jup  & Sat  & -0.006$\pm$0.060   & 0.738$\pm$ $10^{-8}$ &  0.73$\pm$0.060  \\
8 & Mer  & Ura  & -0.006$\pm$0.152   & 0.142$\pm$ $10^{-8}$ & 0.146$\pm$0.152  \\
9 & Ear  & Ura  & 0.0003$\pm$0.0087  & 0.228$\pm$ $10^{-9}$ &  0.21$\pm$0.0087  \\
10& Mar  & Ura  & 0.0002$\pm$0.0048  & 0.281$\pm$ $10^{-8}$ &   0.27$\pm$0.0048 \\
11& Jup  & Ura  & 0.01$\pm$0.31     & 0.52$\pm$  $10^{-9}$  &  0.51$\pm$0.31    \\
\hline
\end{tabular}
\label{res}
\end{table}

\end{document}